\documentclass[aps,pra,superscriptaddress,preprint]{revtex4-1}
\usepackage{subfigure}
\usepackage{xcolor}
\usepackage{graphicx}
\usepackage{tikz}
\usepackage{bm}
\usetikzlibrary{shapes.geometric, arrows, positioning}
\usepackage{amsmath}
\usepackage{float}
\usepackage{ulem}

\usepackage[colorlinks,linkcolor=blue,anchorcolor=blue,citecolor=blue,urlcolor=blue]{hyperref}

\newcommand{\cZ}{\mathcal{Z}}

\begin{document}

		\title{Least constraint approach to non‑relativistic quantum mechanics}
		
		\author{Ning Liu}
		\email{ningliu@mail.bnu.edu.cn}
		\address{School of Physics and Astronomy, Anqing Normal University, Anqing 246133, China}
		\address{Key Laboratory of Multiscale Spin Physics (Ministry of Education), Beijing Normal University, Beijing 100875, China}
		
		\begin{abstract}
			We formulate a variational principle for non‑relativistic quantum mechanics inspired by Gauss's principle of least constraint. We define a quantum constraint functional as the probability‑weighted square deviation between the actual motion and the unconstrained motion that would arise from external forces alone. In this functional, the quantum potential plays the role of an intrinsic constraint that modifies the acceleration. Minimizing this quantum constraint functional with respect to the acceleration field yields the quantum Euler equations, which together with the continuity equation are equivalent to the Schr\"odinger equation. The principle is instantaneous and provides a differential characterization of quantum evolution. We demonstrate that this formulation is not a mere rewriting of existing dynamics: it provides a unified and technically economical treatment of geometric constraints and velocity‑dependent dissipative forces, neither of which admits a straightforward global variational formulation. Potential applications to a broad range of quantum phenomena are also indicated.
		\end{abstract}
		\maketitle
	
	\section{Introduction}
	
	Variational principles play a foundational role in physics. The most familiar forms, such as Hamilton's principle and the Feynman path integral~\cite{feynman1948}, are global in time: they select the actual trajectory from the space of all possible histories over a finite time interval~\cite{rojobloch}. Despite their elegance and power, global variational principles face inherent limitations when applied to systems that do not admit a well‑defined Lagrangian or Hamiltonian. In classical mechanics, prominent examples include systems with nonholonomic constraints, friction, or time‑dependent boundary conditions. In quantum mechanics, similar difficulties arise: dissipative forces cannot be derived from a real potential~\cite{kostin1972}, velocity‑dependent interactions break the standard Legendre transform~\cite{gergely2002}, and the quantization of constrained systems often requires delicate limiting procedures or Dirac‑Bergmann constraint analysis~\cite{dirac1950} to eliminate unphysical degrees of freedom. These difficulties suggest that global variational principles, although deep, are not always the most natural or convenient tool. 
	
	An alternative class of differential variational principles dispenses with the integration over time and instead imposes an extremum condition at each instant. Among them, Gauss's principle of least constraint~\cite{gauss1829} is distinguished by its intuitive physical content: the true motion of a constrained system minimizes, at every moment, the weighted square deviation 
	\begin{equation}
		Z = \sum_{i} m_i \left( \ddot{\bm{r}}_i - \frac{\bm{F}_i}{m_i} \right)^2,
		\label{eq:gauss}
	\end{equation}
	where \(m_i\) are the masses, \(\ddot{\bm{r}}_i\) the actual accelerations, and \(\bm{F}_i\) the given (active) forces. The quantity in parentheses is precisely the acceleration contributed by the constraint forces, so the principle asserts that nature chooses the motion that makes the constraint forces as small as possible in a least‑squares sense.
	
	Because Gauss's principle is local in time, it does not require the existence of a potential function or a global action. Non‑conservative forces (e.g., dissipative forces) can be included by simply adding them to the ``unconstrained'' acceleration term, without modifying the variational structure. Geometric constraints are handled by projecting the allowed accelerations onto the tangent plane of the constraint manifold at each instant. This makes it a remarkably flexible framework for constrained and dissipative dynamics.
	
\begin{figure}[!ht]
		\centering
		\begin{tikzpicture}[
			box/.style={rectangle, rounded corners, minimum width=2.3cm, minimum height=1.3cm, text centered, align=center, draw=black},
			arrow/.style={thick,->,>=stealth}
			]
			
			\node (principle) [box] {principle of\\ least constraint};
			
			\node (quantum) [box, above left=0.6cm and 0.6cm of principle] {Quantum constraint\\ \(\cZ=?\)};
			
			\node (classical) [box, above right=0.6cm and 0.6cm of principle] {Classical constraint\\ \(Z = m \left\| \ddot{\mathbf{r}} - \frac{\mathbf{F}}{m} \right\|^2\)};
			
			\node (schrodinger) [box, below left=0.6cm and 0.6cm of principle] {Schr\"odinger's equation\\ \(i\hbar\partial_t\psi = H\psi\)};
			
			\node (newton) [box, below right=0.6cm and 0.6cm of principle] {Newton's equation\\ \(\mathbf{F} = m\ddot{\mathbf{r}}\)};
			
			\draw[arrow] (quantum) -- (principle);
			\draw[arrow] (classical) -- (principle);
			\draw[arrow] (principle) -- (schrodinger);
			\draw[arrow] (principle) -- (newton);
			\draw[arrow] (quantum) -- node[above] {} (classical);
			\draw[arrow] (schrodinger) -- node[above] {} (newton);
		\end{tikzpicture}
		\caption{Schematic of the research motivation.}
		\label{fig:motive}
	\end{figure}
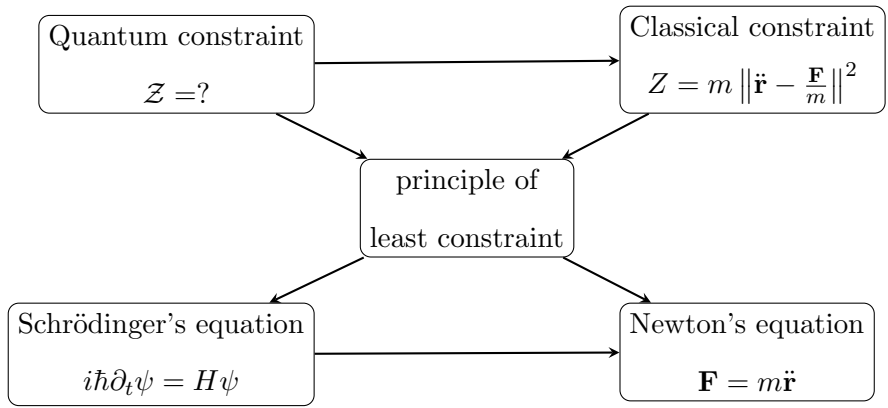

	Given these advantages, it is natural to ask whether an analogous formulation exists for quantum mechanics, as shown in Fig.~\ref{fig:motive}. In this paper, we propose a quantum version of Gauss's principle of least constraint. Working within the Madelung hydrodynamic representation, we introduce an instantaneous quantum constraint functional \(\cZ\) that measures the local deviation between the actual acceleration of the probability fluid and the acceleration due to the combined external and quantum potentials. Minimizing \(\cZ\) with respect to the acceleration field yields the quantum Euler equations; together with the continuity equation, they are shown to be equivalent to the Schr\"odinger equation via the inverse Madelung transformation.
	
	It is crucial to emphasize that this construction is not a mere rewriting of the Schr\"odinger dynamics. The quantum Gauss principle offers two significant advantages that are absent in Lagrangian‑based formulations. First, it is a genuine instantaneous variational principle operating on the acceleration field, and hence can directly incorporate velocity‑dependent forces and geometric constraints without requiring a global potential or Lagrangian. Second, the structure of the constraint functional reveals an intrinsic geometric connection: the quantum potential plays the role of a curvature term arising from the shape of the probability density, so that free quantum evolution minimizes an acceleration mismatch in a manner analogous to Hertz's principle of least curvature~\cite{hertz1894}. These features make the framework particularly transparent and versatile. To demonstrate the effectiveness of this framework, we apply it to two physically important scenarios: a quantum particle constrained to a curved surface (Sec.~\ref{sec:sphere}) and a damped quantum harmonic oscillator (Sec.~\ref{sec:damped}). The first example reveals the dynamical origin of the well‑known geometric potential, while the second yields a rigorous instantaneous variational foundation for the Kostin nonlinear Schr\"odinger equation. Both cases showcase how the least constraint principle unifies constrained and dissipative dynamics within a single variational structure.
	
	The remainder of this paper is organized as follows. Sec.~\ref{sec:constraint} presents the mathematical formulation of the quantum least constraint principle and proves its equivalence to the Schr\"odinger equation. Sec.~\ref{sec:sphere} applies the principle to a particle confined to a sphere, uncovering the geometric potential as a natural force projection. Sec.~\ref{sec:damped} treats the damped harmonic oscillator. Sec.~\ref{sec:conclusion} concludes with a discussion and outlook.
	
	\section{Quantum principle of least constraint}
	\label{sec:constraint}
	
	We consider a non‑relativistic particle of mass \(m\) in an external potential \(V(\mathbf{r})\). The statistical state of the Madelung fluid is described by a probability density \(\rho(\mathbf{r},t)\) and a velocity field \(\mathbf{v}(\mathbf{r},t)\). As in the classical Gauss principle, our aim is to determine the instantaneous acceleration field from the given state.
	
	\subsection{The instantaneous state and the constraint functional}
	
	At any fixed instant \(t\), the fields \(\rho\) and \(\mathbf{v}\) constitute the complete kinematical description of the system, analogous to positions and velocities in particle mechanics. They are prescribed and must satisfy the continuity equation
	\begin{equation}
		\frac{\partial\rho}{\partial t} + \nabla\cdot(\rho\mathbf{v}) = 0 \label{eq:continuity}
	\end{equation}
	as a kinematic constraint ensuring probability conservation. The quantum potential is defined solely in terms of the given density~\cite{bohm1952}:
	\begin{equation}
		Q[\rho] = -\frac{\hbar^2}{2m}\frac{\nabla^2\sqrt{\rho}}{\sqrt{\rho}} .
	\end{equation}
	It acts as an additional potential that encodes the non‑local influence of quantum fluctuations.
	
	The true acceleration of a fluid element is the material derivative
	\begin{equation}
		\frac{D\mathbf{v}}{Dt} = \frac{\partial\mathbf{v}}{\partial t} + (\mathbf{v}\cdot\nabla)\mathbf{v}.
	\end{equation}
	In this expression, the convective term \((\mathbf{v}\cdot\nabla)\mathbf{v}\) is fully determined by the given instantaneous velocity field \(\mathbf{v}\). The only quantity that is unknown and represents the genuine dynamical response of the system is the local time derivative \(\partial\mathbf{v}/\partial t\). Therefore, following the spirit of Gauss's principle, we treat \(\partial\mathbf{v}/\partial t\) as the primary variational variable, while \(\rho\) and \(\mathbf{v}\) are held fixed.
	
	Had the system been a classical fluid without quantum effects, the acceleration would have been \(-(1/m)\nabla V\). The quantum potential introduces an additional contribution, so the ``unconstrained'' acceleration (the one that would occur if the fluid could move solely under the influence of the internal and external forces) is \(-(1/m)\nabla(V+Q)\). We define the quantum constraint functional as
	\begin{equation}
		\cZ\left[\frac{\partial\mathbf{v}}{\partial t}\right] = \int \rho(\mathbf{r}) \left\| \frac{D\mathbf{v}}{Dt} + \frac{1}{m}\nabla V + \frac{1}{m}\nabla Q \right\|^2 d^3r,
		\label{eq:Zdef}
	\end{equation}
	with the understanding that \(\rho\) and \(\mathbf{v}\) are fixed parameters. The norm \(\|\cdot\|\) is the standard Euclidean norm; any constant metric would give equivalent results. We now state the quantum principle of least constraint:
	
	\begin{quote}
		At each instant, subject to the probability‑conservation constraint, the acceleration field is such that the deviation from the unconstrained motion is minimized in the probability‑weighted least‑squares sense. That is, the true motion minimizes \(\cZ\).
	\end{quote}

	\subsection{Equivalence to the Schr\"odinger equation}
	
	The condition \(\delta\cZ = 0\) provides the equation of motion for the fluid. We now show that, together with the continuity equation, it is fully equivalent to the Schr\"odinger equation. 
	
	Varying \(\cZ\) with respect to the independent variable \(\partial\mathbf{v}/\partial t\) yields
	\begin{equation}
		\delta\cZ = \int 2\rho(\mathbf{r}) \left( \frac{D\mathbf{v}}{Dt} + \frac{1}{m}\nabla V + \frac{1}{m}\nabla Q \right) \cdot \delta\left(\frac{\partial\mathbf{v}}{\partial t}\right) d^3r .
	\end{equation}
	Because the variation \(\delta(\partial\mathbf{v}/\partial t)\) is arbitrary and the factor \(\rho(\mathbf{r})\) is non‑negative (and positive on the support of the fluid), the stationary condition \(\delta\cZ = 0\) implies
	\begin{equation}
		\frac{D\mathbf{v}}{Dt} = -\frac{1}{m}\nabla(V+Q). \label{eq:euler3D}
	\end{equation}
	This is the quantum Euler equation. Together with the continuity equation, Eq.~\eqref{eq:continuity}, it forms a closed hydrodynamic system for \(\rho\) and \(\mathbf{v}\).
	
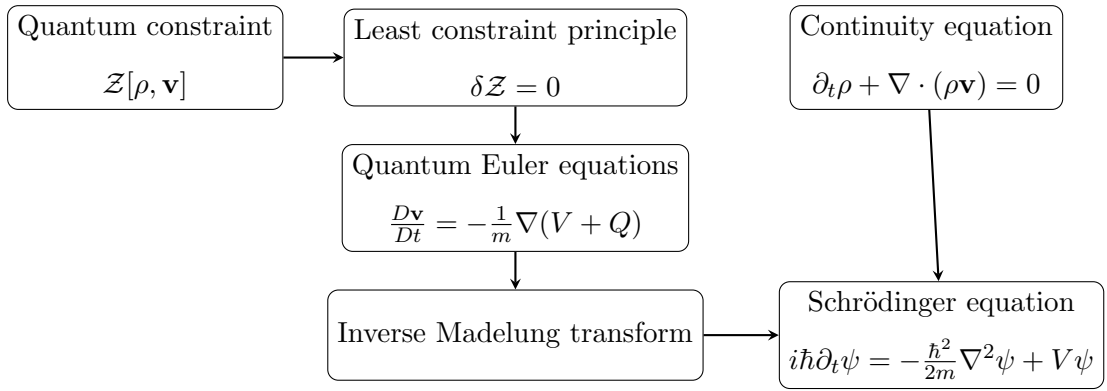
\begin{figure}[!ht]
		\centering
		\begin{tikzpicture}[
			box/.style={rectangle, rounded corners, minimum width=2.2cm, minimum height=1.2cm, text centered, align=center, draw=black},
			arrow/.style={thick,->,>=stealth}
			]
			
			\node (constraint) [box] {Quantum constraint\\ \(\cZ[\rho,\mathbf{v}]\)};
			\node (principle) [box, right=0.8cm of constraint] {Least constraint principle\\ \(\delta\cZ=0\)};
			\node (continuity) [box,  right=1.36cm of principle] {Continuity equation\\ \(\partial_t\rho + \nabla\cdot(\rho\mathbf{v})=0\)};
			
			\node (euler) [box,below=0.5cm of principle] {Quantum Euler equations\\ \(\frac{D\mathbf{v}}{Dt}=-\frac{1}{m}\nabla(V+Q)\)};
			
			\node (transform) [box, below=0.5cm of euler] {Inverse Madelung transform};
			\node (schrodinger) [box, right=1cm of transform] {Schr\"odinger equation\\ \(i\hbar\partial_t\psi=-\frac{\hbar^2}{2m}\nabla^2\psi+V\psi\)};
			
			\draw[arrow] (constraint) -- (principle);
			\draw[arrow] (principle) -- (euler);
			\draw[arrow] (euler) -- (transform);
			\draw[arrow] (continuity) -- (schrodinger);
			\draw[arrow] (transform) -- (schrodinger);
			
		\end{tikzpicture}
		\caption{Flow diagram of the quantum Gauss principle. Minimization of \(\cZ\) yields the quantum Euler equations, which together with the continuity equation lead to the Schr\"odinger equation via the inverse Madelung transformation.}
		\label{fig:derivation}
	\end{figure}

	We now demonstrate that Eqs.~\eqref{eq:continuity} and \eqref{eq:euler3D} are equivalent to the Schr\"odinger equation. Assume the velocity field is irrotational, \(\mathbf{v} = \nabla S/m\). The continuity equation becomes
	\begin{equation}
		\frac{\partial\rho}{\partial t} + \frac{1}{m}\nabla\cdot(\rho\nabla S) = 0,
	\end{equation}
	which, for \(\rho>0\), can be written as
	\begin{equation}
		\frac{\partial}{\partial t}\ln\rho + \frac{1}{m}\nabla S \cdot \nabla\ln\rho + \frac{1}{m}\nabla^2 S = 0. \label{eq:continuity_log}
	\end{equation}
	The quantum Euler equation is the gradient of the quantum Hamilton–Jacobi equation,
	\begin{equation}
		\frac{\partial S}{\partial t} + \frac{1}{2m}|\nabla S|^2 + V - \frac{\hbar^2}{2m}\frac{\nabla^2\sqrt{\rho}}{\sqrt{\rho}} = 0. \label{eq:QHJ_again}
	\end{equation}
	Using the identity
	\begin{equation}
		\frac{\nabla^2\sqrt{\rho}}{\sqrt{\rho}} = \frac{1}{2}\nabla^2\ln\rho + \frac{1}{4}|\nabla\ln\rho|^2,
	\end{equation}
	we can combine the two real equations into a single complex equation. To this end, introduce the complex field
	\begin{equation}
		\Phi = \frac{1}{2}\ln\rho + \frac{i}{\hbar}S .
	\end{equation}
	Multiplying its time derivative by \(i\hbar\) gives
	\begin{equation}
		i\hbar\frac{\partial\Phi}{\partial t} = \frac{i\hbar}{2}\frac{\partial}{\partial t}\ln\rho - \frac{\partial S}{\partial t}. \label{eq:step1}
	\end{equation}
	Now replace \(\partial_t\ln\rho\) and \(\partial_t S\) by the spatial derivatives from Eqs.~\eqref{eq:continuity_log} and~\eqref{eq:QHJ_again}. After substitution we obtain
	\begin{equation}
		i\hbar\frac{\partial\Phi}{\partial t} = V + \frac{1}{2m}|\nabla S|^2 - \frac{\hbar^2}{4m}\nabla^2\ln\rho - \frac{\hbar^2}{8m}|\nabla\ln\rho|^2 - \frac{i\hbar}{2m}\nabla^2 S - \frac{i\hbar}{2m}\nabla S\cdot\nabla\ln\rho. \label{eq:step2}
	\end{equation}
	Next, we compute the spatial derivatives of \(\Phi\):
	\begin{equation}
		\nabla\Phi = \frac12\nabla\ln\rho + \frac{i}{\hbar}\nabla S,\qquad 
		\nabla^2\Phi = \frac12\nabla^2\ln\rho + \frac{i}{\hbar}\nabla^2 S,
	\end{equation}
	and the direct square of the gradient
	\begin{align}
		(\nabla\Phi)^2 = \frac14|\nabla\ln\rho|^2 + \frac{i}{\hbar}\nabla\ln\rho\cdot\nabla S - \frac{1}{\hbar^2}|\nabla S|^2. \label{eq:grad_square}
	\end{align}
	Now consider the combination \(-\frac{\hbar^2}{2m}\bigl(\nabla^2\Phi + (\nabla\Phi)^2\bigr) + V\). Inserting the expressions for \(\nabla^2\Phi\) and \((\nabla\Phi)^2\) we find
	\begin{align}
		-\frac{\hbar^2}{2m}\Bigl(\nabla^2\Phi + (\nabla\Phi)^2\Bigr) + V
		&= V - \frac{\hbar^2}{4m}\nabla^2\ln\rho - \frac{i\hbar}{2m}\nabla^2 S - \frac{\hbar^2}{8m}|\nabla\ln\rho|^2 - \frac{i\hbar}{2m}\nabla\ln\rho\cdot\nabla S + \frac{1}{2m}|\nabla S|^2 .
	\end{align}
	This is precisely the right‑hand side of Eq.~\eqref{eq:step2}. Therefore we arrive at the compact nonlinear equation for \(\Phi\):
	\begin{equation}
		i\hbar\frac{\partial\Phi}{\partial t} = -\frac{\hbar^2}{2m}\Bigl( \nabla^2\Phi + (\nabla\Phi)^2 \Bigr) + V. \label{eq:Phi_eq}
	\end{equation}
	Finally, define \(\psi = e^{\Phi} = \sqrt{\rho}\,e^{iS/\hbar}\). Using \(\nabla\psi = \psi\,\nabla\Phi\) and \(\nabla^2\psi = \psi\bigl(\nabla^2\Phi + (\nabla\Phi)^2\bigr)\), Eq.~\eqref{eq:Phi_eq} becomes
	\begin{equation}
		i\hbar\frac{\partial\psi}{\partial t} = -\frac{\hbar^2}{2m}\nabla^2\psi + V\psi. \label{eq:schrodinger_final}
	\end{equation}
	Thus the quantum Gauss principle, supplemented by the continuity equation, reproduces exactly the Schr\"odinger equation. The derivation reveals that the Madelung ansatz \(\psi = \sqrt\rho e^{iS/\hbar}\) is not an ad hoc substitution but rather the linearizing transformation of the underlying hydrodynamic system: the nonlinear complex equation for \(\Phi\) is precisely the logarithmic form of the linear Schr\"odinger equation for \(\psi\). The flow diagram of the logical steps is shown in Fig.~\ref{fig:derivation}.
	
	It is noted that the mapping \(\psi = \sqrt{\rho}e^{iS/\hbar}\) assumes a single‑valued phase \(S\), which is valid on each smooth phase branch. Near caustics or strong interference, the phase becomes multi‑valued and the single velocity field \(\mathbf{v} = \nabla S/m\) is insufficient; the full linear Schr\"odinger equation, Eq.~\eqref{eq:schrodinger_final}, remains globally valid. The quantum Gauss principle, as presented here, should be understood as a variational characterization on each smooth branch. Its extension to multi‑valued flows will be studied elsewhere.
	
	\subsection{Geometric interpretation: acceleration driven by the curvature of the probability density}
	\label{sec:geometric}
	
	When the external potential vanishes, the quantum constraint functional, Eq.~\eqref{eq:Zdef}, reduces to
	\begin{equation}
		\cZ_{\text{free}}\left[\frac{\partial\mathbf{v}}{\partial t}\right] = \int \rho(\mathbf{r})\,
		\left\| \frac{D\mathbf{v}}{Dt} + \frac{1}{m}\nabla Q \right\|^{2} d^{3}r,
	\end{equation}
	and the stationarity condition \(\delta\cZ_{\text{free}}=0\) yields the free quantum Euler equation
	\begin{equation}
		\frac{D\mathbf{v}}{Dt} = -\frac{1}{m}\nabla Q. \label{eq:free_euler}
	\end{equation}
	This equation has a clear geometric meaning: the acceleration of each fluid element is completely determined by the gradient of the quantum potential, which in turn is built entirely from the instantaneous shape of the probability density.
	
	To see this more clearly, introduce the real amplitude \(R(\mathbf{r},t) = \sqrt{\rho(\mathbf{r},t)} > 0\). The quantum potential then reads
	\begin{equation}
		Q = -\frac{\hbar^2}{2m} \frac{\nabla^2 R}{R},
	\end{equation}
	and its gradient is
	\begin{equation}
		\nabla Q = -\frac{\hbar^2}{2m} \, \nabla\!\left(\frac{\nabla^2 R}{R}\right).
	\end{equation}
	The quantity \(\nabla^2 R / R\) is a purely spatial object: it measures the local ``curvature'' of the amplitude function \(R\). In regions where \(R\) bends strongly (e.g., near nodes of the wave function or at the boundary of a wave packet), this curvature becomes large. Eq.~\eqref{eq:free_euler} therefore states that each material point of the probability fluid experiences an acceleration precisely determined by the gradient of this amplitude curvature.
	
	Consequently, the flow lines of the free Madelung fluid are not straight; their instantaneous curvature radius is dictated by the spatial variation of the density profile. In classical mechanics, Hertz's principle of least curvature~\cite{hertz1894} states that a free system (no active forces) moves with zero acceleration, i.e., along a path of minimal curvature in configuration space. The present quantum analogue possesses the same formal structure: the functional \(\cZ_{\text{free}}\) measures the weighted mean‑square deviation between the true material acceleration \(D\mathbf{v}/Dt\) and the ``intrinsic reference acceleration'' \(-\nabla Q/m\) provided by the density shape. By requiring this deviation to be as small as possible—indeed zero—the principle selects the unique acceleration field that matches the gradient of the amplitude curvature at every point. In this precise sense, the quantum Gauss principle is a natural extension of the classical Hertz principle: nature still minimizes the acceleration mismatch, but the reference acceleration is no longer zero; it is the effective acceleration impressed by the quantum potential.
	
	This geometric interpretation also clarifies why free quantum evolution is non‑dispersive in a nontrivial way. If the acceleration were zero, the continuity equation would simply advect the initial density with a steady velocity field, and the wave packet would not spread. The inevitable spreading of a free quantum wave packet is thus a direct manifestation of the non‑vanishing gradient \(\nabla Q\), which continuously reshapes the velocity field according to the instantaneous curvature of the density profile. In the classical limit \(\hbar \to 0\), the quantum potential vanishes and one recovers the familiar Hertz principle: the squared deviation reduces to \(\int \rho \|D\mathbf{v}/Dt\|^2\), whose minimum is attained at \(D\mathbf{v}/Dt = 0\), i.e., uniform rectilinear motion for each fluid element.
	
	\section{Quantum particle constrained to a sphere}
	\label{sec:sphere}
	
	The confinement of a quantum particle to a curved surface is a physically well-motivated problem, appearing, for instance, in the description of electrons on nanostructured interfaces. It is well understood that a constraint in quantum mechanics cannot be imposed in the classical sense but must be treated as a limiting process: a strong confining potential localizes the particle in an increasingly thin layer around the surface. The pioneering work of da Costa~\cite{daCosta1981} established this thin-layer limit starting from the three-dimensional Schr\"odinger equation. By adopting curvilinear coordinates adapted to the surface, separating normal and tangential degrees of freedom, and taking the squeezing parameter to infinity, one arrives at an effective Schr\"odinger equation on the surface. The remarkable feature of this effective theory is the emergence of a geometric potential \(V_s = -\frac{\hbar^2}{2m}(M^2-K)\), which depends on the surface's mean curvature \(M\) and Gaussian curvature \(K\). This potential is a consequence of the embedding and cannot be derived from the surface's intrinsic metric alone~\cite{daCosta1982}.
	
	While rigorous, the conventional thin-layer method involves lengthy algebraic manipulations and operator-ordering subtleties. The quantum Gauss principle offers not merely a more economical derivation, but more importantly, a direct insight into the dynamical origin of this geometric potential. It shifts the perspective from coordinate space to acceleration space, revealing the geometric force as a necessary consequence of least constraint physics.
	
	We consider a particle of mass \(m\) confined near the surface of a sphere of radius \(R\). The constraint is enforced by a strong harmonic potential \(V_\lambda(q) = \frac{1}{2} m\lambda^{2} q^{2}\), where \(\lambda\to\infty\), localizing the particle in the normal direction. We adopt the coordinates \((\theta,\phi,q)\), where the position vector is \(\mathbf{R}(\theta,\phi,q) = (R+q)\,\hat{\mathbf{r}}(\theta,\phi)\). The three-dimensional metric is diagonal: \(G_{\theta\theta}=(R+q)^2\), \(G_{\phi\phi}=(R+q)^2\sin^2\theta\), \(G_{qq}=1\).
	
	The key insight of the quantum Gauss principle is to formulate the constraint directly in acceleration space. The unconstrained reference acceleration is set by the combined action of the external and quantum potentials, \(-\frac{1}{m}\nabla(V_\lambda + Q)\). The true acceleration \(D\mathbf{v}/Dt\) is, however, forced by the constraint to lie within the tangent bundle of the sphere. The principle posits that nature selects the unique acceleration field that minimizes the probability-weighted deviation from this reference, the quantum constraint functional
	\begin{equation}
		\cZ = \int \rho\left\| \frac{D\mathbf{v}}{Dt} + \frac{1}{m}\nabla V_\lambda + \frac{1}{m}\nabla Q \right\|^2 dV,
	\end{equation}
	with \(dV = (R+q)^2\sin\theta\,d\theta d\phi dq\).
	
	In the limit \(\lambda\to\infty\), the normal motion becomes frozen. The probability density factorizes as \(\rho \approx \rho_t(\theta,\phi)\,|\psi_n(q)|^2\), and the normal velocity vanishes, \(v_n = 0\). After integrating over the normal degree of freedom, the constraint functional reduces to a purely tangential one,
	\begin{equation}
		\cZ_{\text{eff}} = \int \rho_t\left\| \frac{D\mathbf{v}_t}{Dt} + \frac{1}{m}\nabla_S Q_t \right\|_g^2 dS.
	\end{equation}
	A crucial step is the projection of the original three-dimensional quantum constraint onto the tangent bundle. The normal component of the unconstrained reference acceleration cannot be satisfied by the true motion and acts back as an effective tangential force. This projection naturally generates the geometric potential, which is given by the familiar da Costa formula~\cite{daCosta1981}
	\begin{equation}
		V_s = -\frac{\hbar^2}{2m}\bigl(M^2 - K\bigr).
	\end{equation}
	Thus, the functional takes the form
	\begin{equation}
		\cZ_{\text{eff}} = \int \rho_t\left\| \frac{D\mathbf{v}_t}{Dt} + \frac{1}{m}\nabla_S V_s + \frac{1}{m}\nabla_S Q_t \right\|_g^2 dS,
	\end{equation}
	where \(dS = R^2\sin\theta\,d\theta d\phi\), \(\nabla_S\) is the covariant derivative on the sphere, \(\|\cdot\|_g\) uses the sphere's metric, and \(Q_t = -\frac{\hbar^2}{2m}\frac{\Delta_S\sqrt{\rho_t}}{\sqrt{\rho_t}}\) with \(\Delta_S\) the Laplace–Beltrami operator.
	
	For the sphere, the curvatures are constant: \(M=1/R\), \(K=1/R^2\), so the geometric potential vanishes, \(V_s=0\). In this special case, the constraint force is completely encoded in the kinematic restriction to the tangent space, and we recover free motion on the sphere. For arbitrary surfaces, \(V_s\) is non-zero and directly influences the effective dynamics.
	
	Varying \(\cZ_{\text{eff}}\) with respect to \(\partial\mathbf{v}_t/\partial t\) yields the quantum Euler equation on the sphere,
	\begin{equation}
		\frac{D\mathbf{v}_t}{Dt} = -\frac{1}{m}\nabla_S(V_s + Q_t).
	\end{equation}
	Combined with the continuity equation on the sphere, the inverse Madelung transformation yields precisely the effective Schr\"odinger equation first obtained by da Costa~\cite{daCosta1981},
	\begin{equation}
		i\hbar\frac{\partial\psi_t}{\partial t} = -\frac{\hbar^2}{2m}\Delta_S\psi_t + V_s\,\psi_t.
	\end{equation}
	
	This equivalence is instructive. In the conventional derivation, \(V_s\) appears to emerge from operator-ordering subtleties when expanding the Laplacian in curvilinear coordinates. In the quantum Gauss principle, it appears as a force-term projection from the full acceleration space onto the allowed tangent space. The geometric potential is thus not a mathematical artifact, but a dynamical consequence of a fundamental tension: the quantum system, driven by the quantum potential \(Q\), has a natural tendency to evolve in a manner incompatible with the geometric constraint. The least constraint principle selects the unique allowed motion that best approximates this unconstrained evolution, and the geometric potential is a quantitative measure of the irreducible mismatch between the two.
	
	This perspective offers both conceptual clarity and technical economy. The constraint is applied directly to the acceleration field, bypassing the cumbersome coordinate expansions of the thin-layer limit. Furthermore, the method generalizes immediately to any surface defined by a level set \(f(\mathbf{r})=0\), by employing a confining potential \(V_\lambda(\mathbf{r}) = \frac12 m\lambda^2 f(\mathbf{r})^2\). For such cases, the explicit change to curvilinear coordinates and the subsequent expansion of the Laplacian become prohibitively complex, yet the projection onto the acceleration space remains a well-defined local operation.
	
	\section{Damped quantum harmonic oscillator}
	\label{sec:damped}
	
	Dissipative quantum systems are often described by effective nonlinear Schr\"odinger equations. A prominent example is the Schr\"odinger–Langevin equation introduced by Kostin~\cite{kostin1972}, which incorporates a logarithmic nonlinearity to ensure exponential decay of the mean momentum while preserving probability conservation. The quantum Gauss principle provides an instantaneous variational foundation for this equation, without the need to postulate the damping law at the level of expectation values.
	
	Consider a one‑dimensional harmonic oscillator with potential \(V(x) = \frac12 m\omega_0^2 x^2\) and a linear damping force \(-\gamma v\) (\(\gamma>0\)). In the Madelung representation (Sec.~\ref{sec:constraint}), the velocity field is \(v = \partial_x S / m\). The classical equation of motion for a representative particle would be \(m\ddot{x} = -V'(x) - \gamma \dot{x}\), which corresponds to an ``unconstrained'' acceleration \(-\frac1m V' - \frac{\gamma}{m}v\). Following the spirit of the Gauss principle, we include this velocity‑dependent term directly in the reference acceleration, together with the quantum potential. The extended quantum constraint functional therefore reads
	\begin{equation}
		\cZ_{\text{diss}}\!\left[\frac{\partial v}{\partial t}\right]
		= \int \rho(x) \left( \frac{Dv}{Dt} + \frac{1}{m}V' + \frac{\gamma}{m}v + \frac{1}{m}Q' \right)^{\!2} dx,
		\label{eq:Zdamp}
	\end{equation}
	where the material derivative \(Dv/Dt\) is the one‑dimensional restriction of the expression given in Sec.~\ref{sec:constraint}. Varying \(\cZ_{\text{diss}}\) with respect to \(\partial v/\partial t\) (with \(\rho\) and \(v\) held fixed) yields the damped quantum Euler equation
	\begin{equation}
		\frac{Dv}{Dt} = -\frac{1}{m}V' - \frac{\gamma}{m}v - \frac{1}{m}Q'.
		\label{eq:damped_euler}
	\end{equation}
	
	To obtain the corresponding wave‑function equation we employ the Madelung decomposition \(\psi = \sqrt{\rho}\,e^{iS/\hbar}\). Substituting \(v = \partial_x S/m\) and using the identity \(Dv/Dt = \frac1m \partial_x\bigl(\partial_t S + \frac{1}{2m}(\partial_x S)^2\bigr)\), Eq.~\eqref{eq:damped_euler} becomes
	\begin{equation}
		\frac{\partial}{\partial x}\!\left( \frac{\partial S}{\partial t} + \frac{1}{2m}(\partial_x S)^2 + V + Q \right)
		= -\frac{\gamma}{m}\,\partial_x S .
	\end{equation}
	Integrating with respect to \(x\) and absorbing the integration constant into the phase (a freedom that leaves the physical velocity and density unchanged) we obtain the damped quantum Hamilton–Jacobi equation
	\begin{equation}
		\frac{\partial S}{\partial t} + \frac{1}{2m}(\partial_x S)^2 + V + Q = -\frac{\gamma}{m}\,S .
		\label{eq:dampedQHJ}
	\end{equation}
	Eq.~\eqref{eq:dampedQHJ} differs from the standard quantum Hamilton–Jacobi equation only by the dissipative term \(-\gamma S/m\). Using the relation \(\ln(\psi/\psi^*) = (2i/\hbar) S\), which follows directly from the Madelung ansatz, one finds \(S = (\hbar/2i)\ln(\psi/\psi^*)\). Inserting this into \eqref{eq:dampedQHJ} and comparing with the general Madelung form \(\partial_t S + \frac{1}{2m}(\partial_x S)^2 + V_{\text{eff}} + Q = 0\) shows that the effective potential acquires an imaginary part,
	\begin{equation}
		V_{\text{eff}} = V(x) + \frac{i\hbar\gamma}{2m}\ln\frac{\psi}{\psi^*} .
		\label{eq:Veff}
	\end{equation}
	The wavefunction therefore satisfies the nonlinear Schr\"odinger equation
	\begin{equation}
		i\hbar\frac{\partial\psi}{\partial t}
		= \left( -\frac{\hbar^2}{2m}\frac{\partial^2}{\partial x^2}
		+ \frac12 m\omega_0^2 x^2
		+ \frac{i\hbar\gamma}{2m}\ln\frac{\psi}{\psi^*} \right)\psi,
		\label{eq:kostin}
	\end{equation}
	which is precisely the Kostin equation~\cite{kostin1972}. The logarithmic term is purely imaginary and therefore does not disturb the continuity equation, while guaranteeing that the expectation value of the momentum decays exponentially, \(\frac{d}{dt}\langle P \rangle = -\frac{\gamma}{m}\langle P \rangle - \langle V' \rangle\). In the classical limit \(\hbar\to0\) the quantum potential disappears and Eq.~\eqref{eq:dampedQHJ} reduces to the classical damped Hamilton–Jacobi equation, leading to Newton's equation \(m\ddot{x} = -m\omega_0^2 x - \gamma \dot{x}\).
	
	This derivation demonstrates that the quantum Gauss principle unifies conservative and dissipative dynamics within a single variational framework. The damping law is encoded in the unconstrained acceleration, and the variational principle automatically generates the nonlinearity required in the wave‑function representation.
	
	\section{Conclusion}
	\label{sec:conclusion}
	
	We have formulated a quantum version of Gauss's principle of least constraint for non‑relativistic quantum mechanics and demonstrated its equivalence to the Schr\"odinger equation. The principle defines an instantaneous quantum constraint functional whose minimization yields the quantum Euler equations, supplemented by the continuity equation. We have shown that this approach is not a trivial rewriting: it provides a differential variational framework that naturally handles geometric constraints and velocity‑dependent dissipative forces. Two concrete applications—a particle on a sphere and a damped harmonic oscillator—illustrate these advantages. The sphere example reveals that the well‑known geometric potential arises from a force projection, a dynamical necessity to reconcile the quantum tendency to accelerate (driven by the quantum potential) with the geometric confinement. The damped oscillator example gives an instantaneous variational foundation to the Kostin equation, demonstrating that non‑conservative forces can be incorporated without altering the variational structure.
	
	A detailed geometric analysis further reveals that free quantum evolution is driven by the curvature of the probability density, establishing a direct structural analogy with Hertz's principle of least curvature~\cite{hertz1894} and clarifying the dynamical role of the quantum potential. This perspective opens several avenues for future work. The extension to multi‑valued flows (e.g., near caustics) is essential for a complete description of strong interference phenomena. The treatment of nonholonomic constraints and many‑particle systems, as well as the extension to quantum field theories, also offer promising directions. Finally, the local variational structure of \(\cZ\) suggests its potential as a tool for studying a variety of quantum phenomena, including tunneling and the hydrodynamic description of Bose–Einstein condensates; these investigations will be pursued elsewhere.
	
	\paragraph*{Acknowledgments}
	This work was supported by the Open Fund of Key Laboratory of Multiscale Spin Physics (Ministry of Education), Beijing Normal University (Grant No. SPIN2024N03), and the Scientific Research Startup Foundation for High‑Level Talents at Anqing Normal University (Grant No. 241042). The author thanks Zhanchun Tu, Quanhui Liu and Yukun Yang for valuable discussions and suggestions.


\begin{thebibliography}{99}
		
		\bibitem{feynman1948}
		R.P. Feynman, Space-time approach to non-relativistic quantum mechanics,
		Rev. Mod. Phys. 20 (1948) 367--387.
		\url{https://doi.org/10.1103/RevModPhys.20.367}.
		
		\bibitem{rojobloch}
		A. Rojo, A. Bloch, \emph{The Principle of Least Action: History and Physics},
		Cambridge University Press, Cambridge, 2018.
		
		\bibitem{kostin1972}
		M.D. Kostin, On the Schr\"{o}dinger–Langevin equation,
		J. Chem. Phys. 57 (1972) 3589--3591.
		\url{https://doi.org/10.1063/1.1678812}.
		
		\bibitem{gergely2002}
		L.\'{A}. Gergely, On Hamiltonian formulations of the Schr\"{o}dinger system,
		Ann. Phys. 298 (2002) 394--402.
		\url{https://doi.org/10.1006/aphy.2002.6262}.
		
		\bibitem{dirac1950}
		P.A.M. Dirac, Generalized Hamiltonian dynamics,
		Can. J. Math. 2 (1950) 129--148.
		\url{https://doi.org/10.4153/CJM-1950-012-1}.
		
		\bibitem{gauss1829}
		C.F. Gauss, \"{U}ber ein neues allgemeines Grundgesetz der Mechanik,
		J. Reine Angew. Math. 4 (1829) 232--235.
		
		\bibitem{hertz1894}
		H. Hertz, \emph{Die Prinzipien der Mechanik in neuem Zusammenhange dargestellt}
		(1894); English translation: \emph{The Principles of Mechanics Presented in a New Form},
		Dover, New York, 1956.
		
		\bibitem{bohm1952}
		D. Bohm, A suggested interpretation of the quantum theory in terms of
		``hidden'' variables, I and II,
		Phys. Rev. 85 (1952) 166--179; 180--193.
		\url{https://doi.org/10.1103/PhysRev.85.166}
		\url{https://doi.org/10.1103/PhysRev.85.180}.
		
		
		\bibitem{daCosta1981}
		R.C.T. da Costa, Quantum mechanics of a constrained particle,
		Phys. Rev. A 23 (1981) 1982--1987.
		\url{https://doi.org/10.1103/PhysRevA.23.1982}.
		
		\bibitem{daCosta1982}
		R.C.T. da Costa, Constraints in quantum mechanics,
		Phys. Rev. A 25 (1982) 2893--2900.
		\url{https://doi.org/10.1103/PhysRevA.25.2893}.
		
	\end{thebibliography}
\end{document}